\documentclass[onecolumn,preprintnumbers,elsart]{revtex4}
\usepackage{eurosym}
\usepackage{makeidx}
\usepackage{amssymb}
\usepackage{amsmath}

\begin{document}

\title{Editorial}
\author{Boris A. Malomed$^{1,2}$}
\address{$^{1}$Department of Physical Electronics, School of Electrical
Engineering, Faculty of Engineering, Tel Aviv University, Tel Aviv 69978,
Israel\\
$^{2}$ITMO University, St. Petersburg 197101, Russia}
\maketitle

\section{An overview of the topic and its ramifications}

\subsection{Introduction}

Guided waves represent a vast class of phenomena in which the propagation of
collective excitations in various media is steered in required directions by
fixed (or, sometimes, reconfigurable) conduits. Arguably, the most
well-known and practically important waveguides are single-mode and
multi-mode optical fibers \cite{Crisp-Elliot,Agrawal}, including their more
sophisticated version in the form photonic crystal fibers \cite{Poli}, and
hollow metallic structures transmitting microwave radiation \cite{microwaves}%
. Light\ pipes, in the form of hollow tubes with reflecting inner surfaces,
are used in illumination techniques. On the other hand, medical stethoscopes
offer a commonly known example of a practically important acoustic
waveguide. New directions of studies in photonics are focused on waveguides
for plasmonic waves on metallic surfaces \cite{Hutter,Maier,Bozhevolnyi}
(which provide a possibility of using wavelength much smaller than those
corresponding to the traditional optical range, and thus offer opportunities
to built much more compact photonic devices) and, on the other hand, on
guided transmission of terahertz waves, which also have a great potential
for applications \cite{tera}.

Outside of the realm of photonics (optics and plasmonics) and acoustics,
wave propagation plays a profoundly important role in many other areas, and,
accordingly, waveguiding settings have drawn a great deal of interest in
those areas too. In particular, as concerns hydrodynamics, natural
waveguides, which may be very long, exists for internal waves propagating in
stratified liquids (in fact, in the ocean) \cite{Roger}. Various settings in
the form of waveguides for matter waves are well known in studies of
Bose-Einstein condensates in ultracold bosonic gases \cite{Sengstock,von}.
In solid state physics, guided propagation regimes for magnon waves in
ferromagnetic media are a subject of theoretical and experimental studies
\cite{magnon}. In superconductivity, long Josephson junctions are, as a
matter of fact, waveguides for plasma waves \cite{tera,Josephson}. The
significance of waveguiding in plasma physics is well known too, see, e.g.,
Refs. \cite{Prokhorov,Milchberg,Spence}.

Below, a very brief overview of basic theoretical models and experimental
realizations of various physical implementations of the waveguiding
phenomenology is given. The text is structured according to the character of
the guided wave propagation: linear or nonlinear, and conservative or
dissipative, as well as according to the materials used in the underlying
settings, natural or artificial.

The presentation definitely does not aim to include exhaustive bibliography
on this vast research area. References are given, chiefly, to review
articles and books summarizing the known results, rather than to original
papers where the results were first published. although in some cases
original papers are cited too, if it is necessary in the context of the
presentation.

\subsection{Linear waveguides}

The basic waveguiding structure is a single-mode conduit, designed with a
sufficiently small transverse size and boundary conditions at the boundary
between the guiding core and surrounding cladding, which admits the
propagation of a single transverse mode, while all higher-order modes get
imaginary propagation constants, i.e., they actually cannot propagate. A
commonly known and, arguably, the most important example is provided by
single-mode optical fibers (although, strictly speaking, all such fibers are
bimodal, if the polarization of light is taken into regard) \cite%
{Snyder,Cronin}. Single-mode waveguides are crucially important components
of telecommunication systems, while other applications, such as the delivery
of powerful laser beams for material processing and the creation of complex
spatiotemporal patterns, are best served by multimode conduits \cite%
{multimode-review,multimode}.

Parallel to waveguiding fibers, a subject of many studies in optics are
planar waveguides. In the corresponding models, as well as in their
fiber-optics counterparts, the evolution variable is the propagation
distance, $z$, see Eq. (\ref{SE}) below (this is a common feature of all
guided-wave-propagation settings, not only in optics, but in other physical
realizations of waveguides as well), while the transverse coordinate, $x$,
in the \textit{spatial domain} plays the same role as the reduced-time
variable,
\begin{equation}
\tau \equiv t-V_{\mathrm{gr}}^{-1}z,  \label{tau}
\end{equation}
where $t$ is the time proper, and $V_{\mathrm{gr}}$ is the group velocity of
the carrier wave, plays in the \textit{temporal domain} in fiber optics. The
waveguiding structure in the planar waveguide is represented, roughly
speaking, by a stripe with a locally increased effective refractive index.

Effective equations which model the temporal-domain propagation of optical
waves in fibers, and the spatial-domain propagation in planar waveguides are
similar to each other, taking the form of the linear Schr\"{o}dinger
equation for local amplitude $u$ of the electromagnetic wave, which is
written here in terms of the spatial-domain propagation, and in the scaled
form:
\begin{equation}
i\frac{\partial u}{\partial z}+\frac{1}{2}\frac{\partial ^{2}u}{\partial
x^{2}}-U(x)u=0.  \label{SE}
\end{equation}%
In particular, the aforementioned stripe waveguiding channel is represented
by trapping potential $U(x)$ in Eq. (\ref{SE}), while the second derivative
in Eq. (\ref{SE}) represents the paraxial (weak) transverse diffraction in
the planar waveguide. A ubiquitous form of the potential is%
\begin{equation}
U(x)=-\epsilon ~\mathrm{sech}^{2}(x/l),  \label{trap}
\end{equation}%
where $\epsilon >0$ determines the effective depth of the potential well,
and $l$ determines its width. In the temporal domain, the transverse
coordinate, $x$, is replaced by the above-mentioned temporal variable (\ref%
{tau}). and the diffraction term in Eq. (\ref{SE}) is replaced by $-(\beta
/2)\partial ^{2}u/\partial \tau ^{2},$ where $\beta $ is the coefficient of
the group-velocity dispersion ($\beta >0$ and $\beta <0$ correspond to the
normal and anomalous dispersion, respectively).

Further, the similarity between the wave-propagation equation (\ref{SE}) in
optics and the Schr\"{o}dinger equation in quantum mechanics suggests a
similarity between the guided transmission of waves in the guiding channel
and propagation of real quantum particles in holding channel potentials \cite%
{Exner}. The consideration of the transport of quantum particles in such
channels gives rise to many intriguing peculiarities, such as, in
particular, the consideration of curved guiding channels. In this context,
it is relevant to mention a well-known result which demonstrates a strong
effect of the confinement, imposed by a pipe-shaped potential, on the
character of the effectively one-dimensional mutual scattering of two
quantum particles, which amounts to full reflection of the colliding
particles \cite{Maxim}. This theoretical prediction had suggested the
experimental realization of the concept of the \textit{Tonks-Girardeau gas},
i.e., a gas composed of \textit{hard-core bosons}, which bounce back from
each other when they collide \cite{TG1,TG2}.

A natural generalization of single-channel waveguides is provided by a
\textit{coupler}, which may be considered as a set of two parallel
waveguiding cores, coupled in the transverse direction by tunneling of
guided wave fields steered by each tunnel in the longitudinal direction. The
respective system of coupled equations for amplitudes $u$ and $v$ of
electromagnetic waves in the two cores is \cite{Snyder-JOSA-B} (cf. Eq. (\ref%
{SE}))%
\begin{eqnarray}
i\frac{\partial u}{\partial z}+\frac{1}{2}\frac{\partial ^{2}u}{\partial
x^{2}}+\kappa v-U(x)u &=&0.  \notag \\
&&  \label{uv} \\
i\frac{\partial v}{\partial z}+\frac{1}{2}\frac{\partial ^{2}v}{\partial
x^{2}}+\kappa u-U(x)v &=&0,  \notag
\end{eqnarray}%
where $\kappa $ is the coefficient of the linear inter-core coupling.

The next step is to consider \textit{arrayed systems}, composed of many
parallel guiding cores, which are also coupled in the transverse
direction(s) by the tunneling of longitudinally guided wave fields (planar
and bulk arrays have, respectively, one or two transverse coordinates). The
simplest model of such a guiding medium is provided by the two- or
three-dimensional scaled Schr\"{o}dinger equation with a periodic transverse
potential, which represents the (idealized) structure of the multi-core
bundle:%
\begin{equation}
i\frac{\partial u}{\partial z}+\frac{1}{2}\left( \frac{\partial ^{2}u}{%
\partial x^{2}}+\frac{\partial ^{2}u}{\partial y^{2}}\right) -\epsilon \left[
\cos \left( \frac{2\pi x}{l}\right) +\cos \left( \frac{2\pi y}{l}\right) %
\right] u=0.  \label{lattice}
\end{equation}%
Here $l$ is the array's period (defined in scaled units, in which Eq. (\ref%
{lattice}) is written), and $2\epsilon $ is the scaled depth of the
effective trapping potential. In particular, in optics bulk arrays have been
created, as permanent structures, by burning (also by means of an optical
technology) a large number of parallel guiding cores in a bulk piece of
silica \cite{Jena}. As concerns planar guiding arrays, an interesting
ramification of the topic is the propagation of optical waves in such arrays
made with a curved shape \cite{curved}. On the other hand, a technology for
the creation of reconfigurable \textit{virtual} conduit patterns in the form
of photonic lattices, was elaborated for photorefractive materials \cite%
{Segev}. The latter technology makes use of the fundamental property of the
photorefractive materials, in which the propagation conditions for light
with ordinary and extraordinary polarizations are linear and nonlinear,
respectively. To create a photonic lattice, the experimentalist first
illuminates the sample by counterpropagating pairs of mutually coherent
laser beams in the ordinary polarization, which create a classical
interference pattern in the photorefractive crystal, which is an effectively
linear medium for these beams. Next, a probe beam is launched, with the
extraordinary polarization, in the transverse direction. Due to its inherent
nonlinearity, the probe beam is affected by the originally created photonic
lattice, as if it is a material structure that creates a spatially periodic
modulation of the local refractive index in the transverse directions, i.e.,
essentially, another version of the multi-core guiding structure.

The propagation of light or waves of a different physical nature in arrays
with weak coupling between guiding cores may be naturally approximated by
the discrete Schr\"{o}dinger equation. The basic realization of such a
medium is represented by planar arrays of parallel optical waveguides,
coupled by evanescent waves penetrating dielectric barriers separating
individual cores, the basic model being a scaled discrete version of Eq. (%
\ref{SE}):%
\begin{equation}
i\frac{du_{n}}{dz}+\frac{1}{2}\left( u_{n+1}+u_{n-1}-2u_{n}\right)
-U_{n}u_{n}=0,  \label{discr}
\end{equation}%
where the discrete coordinate, $n$, which replaces $x$, is the number of the
guiding core in the array. The study of light propagation in various
multi-core systems, which may be approximated by lattice models similar to
Eq. (\ref{discr}), is a vast area known as \textit{discrete optics} \cite%
{discr-opt}.

\subsection{Nonlinear waveguides}

In many situations, tightly confined guided waves, propagating in conduits
with a small effective cross-section area, acquire high amplitudes, which is
a source of a great many of fascinating nonlinear effects. In particular,
waveguides often provide a combination of the nonlinearity, group-velocity
dispersion, and low (or, sometimes, completely negligible) losses, which are
necessary ingredients for the creation of \textit{solitons} (robust
self-trapped solitary waves). The simplest and, actually, ubiquitous model
of the nonlinear wave propagation is based on the nonlinear Schr\"{o}dinger
equation (NLSE), which, in the simplest case, includes a cubic term. In
optics, this term represents the Kerr effect, i.e., nonlinear self-focusing
(or, sometimes, self-defocusing) of light in the dielectric medium. The
accordingly amended linear Schr\"{o}dinger equation (\ref{SE}) becomes the
NLSE:
\begin{equation}
i\frac{\partial u}{\partial z}+\frac{1}{2}\frac{\partial ^{2}u}{\partial
x^{2}}-U(x)u+\sigma |u|^{2}u=0,  \label{NLSE}
\end{equation}%
where $\sigma =+1$ and $-1$ corresponds, respectively, to the self-focusing
and defocusing nonlinearity, i.e., self-attraction and self-repulsion of
light in the nonlinear medium. Equations (\ref{uv}) and (\ref{lattice}) each
acquire the same cubic terms as in Eq. (\ref{NLSE}). In particular, the
nonlinear version of Eqs. (\ref{uv}),%
\begin{eqnarray}
i\frac{\partial u}{\partial z}+\frac{1}{2}\frac{\partial ^{2}u}{\partial
x^{2}}+\kappa v-U(x)u+\sigma |u|^{2}u &=&0.  \notag \\
&&  \label{coupler} \\
i\frac{\partial v}{\partial z}+\frac{1}{2}\frac{\partial ^{2}v}{\partial
x^{2}}+\kappa u-U(x)v+\sigma |v|^{2}v &=&0,  \notag
\end{eqnarray}%
is the basic model of nonlinear couplers, their remarkable property being
\textit{spontaneous symmetry breaking} in the case of self-focusing in the
parallel-coupled cores, $\sigma =+1$ \cite{Snyder-JOSA-B,Progr1,book}.

A remarkable property\ of the one-dimensional NLSE in the absence of the
potential ($U=0$ in Eq. (\ref{NLSE})) is that it is an \textit{integrable
equation}, for which a great manifold of exact solutions, including
multi-soliton states, can be produced by means of a mathematical technique
based on the \textit{inverse scattering transform} \cite{IST,IST2,IST3}.
These are bright and dark solitons, in the cases of self-focusing and
defocusing, respectively. In particular, the exact bright-soliton solution
to Eq. (\ref{NLSE}) with $\sigma =+1$ and $U=0$ is%
\begin{equation}
u(x,z)=\eta \exp \left( \frac{i}{2}\left( \eta ^{2}-c^{2}\right)
z+icx\right) ~\mathrm{sech}\left( \eta (x-cz\right) ,  \label{sol}
\end{equation}%
where $\eta $ and $c$ are, respectively, the arbitrary amplitude and
velocity of the soliton (in fact, in the spatial domain, in terms of which
Eq. (\ref{NLSE}) is written, the soliton represents a self-trapped light
beam, and, accordingly, $c$ is not a velocity, but rather a parameter which
determines the tilt of the beam in the $\left( x,z\right) $ plane).

The discrete Schr\"{o}dinger equation (\ref{discr}) also has its natural
nonlinear counterpart, in the form of discrete NLSE,%
\begin{equation}
i\frac{du_{n}}{dz}+\frac{1}{2}\left( u_{n+1}+u_{n-1}-2u_{n}\right)
-U_{n}u_{n}+\sigma |u_{n}|^{2}u_{n}=0,  \label{DNLS}
\end{equation}%
i.e., a discrete version of NLSE\ (\ref{NLSE}). The discrete NLSE gives rise
to discrete solitons and their bound states, which cannot be found in an
exact form, but may be efficiently produced by numerical and approximate
analytical methods \cite{Panos}. The propagation of nonlinear waves in
discrete waveguiding arrays was a subject of numerous theoretical and
experimental works \cite{discr-opt,Tsironis}.

The multidimensional extension of the NLSE\ also has direct realizations in
optics, as well as in the mean-field model of atomic Bose-Einstein
condensates (BECs) \cite{Pit,Pethick}, and in many other areas. In
particular, the spatial-domain light propagation in bulk media is modelled
by the effectively two-dimensional version of Eq. (\ref{NLSE}), with two
transverse coordinates, $\left( x,y\right) $:
\begin{equation}
i\frac{\partial u}{\partial z}+\frac{1}{2}\left( \frac{\partial ^{2}u}{%
\partial x^{2}}+\frac{\partial ^{2}u}{\partial y^{2}}\right) -U(x,y)u+\sigma
|u|^{2}u=0.  \label{2D}
\end{equation}%
Unlike its one-dimensional counterpart (\ref{NLSE}), Eq. (\ref{2D}) in the
free space ($U\left( x,y\right) =0$) is not integrable. It admits formal
soliton solutions, looked for as%
\begin{equation}
u\left( x,y;z\right) =\exp \left( ikz+iS\theta \right) U_{S}(r),  \label{S}
\end{equation}%
in terms of the polar coordinates $\left( r,\theta \right) $ in the $\left(
x,y\right) $ plane, where $k>0$ is a real propagation constant, $S=0,\pm
1,\pm 2\,,$ ..., is an integer \textit{vorticity}, that may be embedded in
the two-dimensional soliton (shaping it as a \textit{vortex ring}), and $%
U_{S}(r)$ is a real radial amplitude function satisfying boundary conditions
$U_{S}(r)\sim \exp \left( -\sqrt{2k}r\right) $ at $r\rightarrow \infty $,
and $U(r)\sim r^{|S|}$ at $r\rightarrow 0$. Solitons (\ref{S}) with $S=0$
are often called \textit{Townes solitons} \cite{Gadi}. However, the Townes
solitons, as well as their vortex counterparts, with $S\neq 0$ in Eq. (\ref%
{S}), are completely unstable, being vulnerable to destruction by the
\textit{critical collapse} (formation of a singularity after a finite
propagation distance) in the case of $S=0$, and by a still stronger
instability which splits vortex rings with $S\neq 0$ \cite{Gadi}.

An important example of nonintegrable one-dimensional system modelling
nonlinear light propagation in optics is the system of coupled-mode
equations which describe the fiber Bragg gratings, i.e., nonlinear optical
fibers with a periodic lattice of local defects permanently written in their
cladding, with a period equal to half the wavelength of light coupled into
this waveguide. The coupled-mode equations govern the evolution of
amplitudes $u$ and $v$ of right- and left-traveling waves, which are
mutually converted (reflected) into each other by the Bragg grating \cite%
{De,Aceves}:%
\begin{eqnarray}
iu_{t}+iu_{x}+\kappa v+\left( \frac{1}{2}|u|^{2}+|v|^{2}\right) u &=&0,
\notag \\
&&  \label{BG} \\
iv_{t}-iv_{x}+\kappa u+\left( \frac{1}{2}|v|^{2}+|u|^{2}\right) v &=&0,
\notag
\end{eqnarray}%
where $\kappa $ is the Bragg-grating reflectivity, and the group velocity of
the light waves in the fiber is scaled to be $1$. This system admits exact
solutions in the form of solitons, but it is not an integrable one. Such
solitons, moving in the fiber Bragg grating as in the waveguide, have been
created in the experiment \cite{FBG}. Roughly, half of the soliton family is
stable, and half unstable.

The use of fiber Bragg gratings operating in the linear regime has grown
into a large industry with many applications, such as sensors, dispersion
compensators, optical buffers, etc. \cite{Kashyap}.

Another fundamentally important nonlinear model for the guided wave
propagation is the one with the quadratic, alias second-harmonic,
nonlinearity, instead of the cubic (Kerr) term in NLSE (\ref{NLSE}). The
model is based on the propagation equations for complex amplitudes $u\left(
x,z\right) $ and $v(x,z)$ of the fundamental and second harmonics \cite%
{Progr2,Buryak}:%
\begin{gather}  \label{chi2}
iu_{z}+\frac{1}{2}u_{xx}+vu^{\ast }=0,  \notag \\
\\
2iv_{z}-qv+\frac{1}{2}v_{xx}+\frac{1}{2}u^{2}=0,  \notag
\end{gather}%
where $q$ is a real mismatch parameter. Equations (\ref{chi2}), although
being a nonintegrable system, also give rise to solitons, which are
generically found in a numerical form. This solitons form a family which is
chiefly stable, with a small instability area \cite{Progr2,Buryak}.

In BEC\ models, Eq. (\ref{NLSE}), with evolution variable $z$ replaced by
(scaled) time, $t$, is called the Gross-Pitaevskii equation, in which the
cubic term represents, in the mean-field approximation, an average effect of
collisions between atoms \cite{Pit,Pethick}. The natural sign of the
collision-induced term corresponds to self-repulsion (self-defocusing),
i.e., $\sigma =-1$ in Eq. (\ref{NLSE}), but, for atomic species such as $%
^{7} $Li, $^{39}$K, and $^{85}$Rb, the sign may be switched to
self-attraction by means of the Feshbach resonance, which is, in turn,
controlled by a magnetic or laser field acting on the experimental setup
\cite{Feshbach}.

Theoretical and experimental work with solitons and other diverse nonlinear
effects (such as the modulational instability \cite{KA} and rogue waves \cite%
{RW1,RW2}, shock waves, separation of immiscible components in binary
systems, kinks and domain walls \cite{Vachaspati}, instantons \cite%
{Rajaraman}, etc.) is a huge research area in many branches of physics \cite%
{Peyrard}, including optics \cite{KA}, matter waves in atomic BECs \cite%
{soliton-BEC} and BECs of quasi-particles (in particular,
excitons-polaritons) \cite{quasi-particles}, plasmas \cite{plasma-nonlinear}%
, ferromagnetic media \cite{ferro}, long Josephson junctions in
superconductivity \cite{JJ}, acoustics \cite{acoustics-nonlinear}, etc. In
many cases, waveguiding settings offer media in which many species of
solitons can be created and/or stabilized, if the solitons do not exist, or
exist but are unstable, in the respective uniform media. Characteristic
examples are various methods elaborated for the stabilization of
three-dimensional spatiotemporal solitons (\textquotedblleft light bullets"
\cite{Silberberg}), which are subject to strong instabilities in both two-
and three-dimensional uniform media \cite{we,me}. It was demonstrated
experimentally that both fundamental spatiotemporal solitons \cite{Jena-fund}
and ones with embedded vorticity \cite{Jena-vort} can be made stable (in
fact, as semi-discrete solitons) in the above-mentioned systems created as
bundles of parallel waveguiding cores in bulk silica samples \cite{Jena}. In
fact, the commonly known stability of temporal optical solitons in nonlinear
fibers \cite{KA} is also an example of the stabilization of a localized mode
which is, strictly speaking, a three-dimensional one, with the self-trapping
in the temporal (longitudinal) direction induced by the nonlinearity, while
the transverse trapping is secured by the fiber's guiding properties, which
are not essentially affected by the nonlinearity. Furthermore, the stability
of matter-wave solitons in cigar-shaped trapping potentials \cite%
{soliton-BEC} is provided by a similar mechanism, in spite of a completely
different physical nature of the latter setting: the longitudinal
self-trapping is induced by the self-attraction of the condensate, due to
attractive interactions between atoms, while the confining potential
prevents spreading of the condensate's wave function in the transverse
directions. Moderate deviation from the effective one-dimensionality
essentially affects the shape of the matter-wave solitons, but still relies
upon the trapping potential, to prevent the collapse of the
three-dimensional self-attractive condensate \cite{Cuevas}.

\subsection{Waveguides built of artificial materials}

The experimental and theoretical results outlined above were obtained in
naturally existing media (and, accordingly, theoretical models of such
media), or in settings produced by straightforward modifications of natural
media, such as the aforementioned multi-core bundled guiding structures
burnt in bulk silica \cite{Jena,Jena-fund,Jena-vort}.

Still natural, but more unusual, optical materials are photonic crystals
(PhCs) \cite{PC} and quasicrystals \cite{quasi,quasi2}, as well as PhC-based
heterostructures and interfaces \cite{hetero}, and PhC fibers \cite%
{PCF,PCF1,PCF2}, i.e., holey fibers in which inner voids form a PhC
structure in the transverse plane. The difference from the traditional
monolithic conduits, which guide light by means of the appropriate
transverse profile of the refractive index, PhCs implement the \textit{%
bandgap-guidance} principle, steering the transmission of different optical
modes according to the spectral bandgap structure, as induced by the
underlying crystalline lattice.

Related to PhC fibers are waveguides built as large-radius hollow fibers,
with a specially designed multi-layer cladding, which, by means of the
Bragg-reflection mechanism (acting in the radial direction), support \textit{%
omniguiding} regime of the transmission of light in such conduits. As a
result, the omniguiding fibers (alias \textit{Bragg fibers}) may provide a
quasi-single regime of the propagation for selected modes, even if the
large-area fiber is a multi-mode one. This is possible due to the fact that
all the modes, except for the selected one,  will be suppressed by strong
losses \cite{Fink}.

It is relevant to mention that another guidance mechanism is possible too,
which makes use of lattice structures similar to those underlying PhCs and
PhC fibers, but, differently from them, these are \textit{nonlinear lattices}
\cite{RMP}, i.e., spatially periodic modulations of the local nonlinearity
coefficient. Naturally, such nonlinear lattices, and their combinations with
the usual linear lattices \cite{HS}, are appropriate for steering nonlinear
modes -- first of all, solitons \cite{RMP,HS}.

Furthermore, a new mechanism (thus far, elaborated theoretically) for guided
transmission of one- and two-dimensional spatial optical solitons, as well
as their matter-wave counterparts in BEC, makes use of a purely
self-defocusing nonlinearity, growing from center to periphery in the $D$%
-dimensional space faster than $r^D$, where $r$ is the radial coordinate
\cite{Barcelona}. This scheme was predicted to stabilize a large number of
diverse self-trapped (soliton-like) modes, both fundamental ones and complex
topologically organized objects, such as three-dimensional \textit{hopfions}
\cite{hopfion}, i.e., vortex rings with internal twist, which carry two
independent topological numbers: the vorticity and the twist.

PhCs and their various modifications may indeed be considered as natural
materials because such structures are found in various animals, accounting
for their coloration \cite{animals}. On the other hand, the recent progress
in photonics has produced remarkable results in the form of artificially
built media, which exhibit completely novel properties, that are not
possible in natural media, a very important example provided by \textit{%
left-handed metamaterials}, featuring negative values of the refractive
index \cite{left,left2}. This property may be used for realization of
fascinating applications, such as superlensing, which breaks the diffraction
limit of imaging \cite{super}, and optical cloaking, lending partial
invisibility to small objects \cite{cloak}. Other well-known examples of
purposely designed artificial optical media with extraordinary properties
include hyperbolic metamaterials, whose tensors of the dielectric
permittivity and/or magnetic permeability feature principal values of
opposite signs \cite{hyperbolic1,hyperbolic2}, planar metasurfaces \cite%
{meta}, epsilon-near-zero materials, in which the refractive index nearly
vanishes \cite{ENZ}, photonic topological insulators \cite%
{insulator,insulator2} (which exemplify the area of \textit{topological
photonics} \cite{topological}), and others. The use of such media opens
numerous possibilities to implement diverse optical effects, including
nonlinear ones \cite{Maimistov} and guided-wave propagation, in forms that
were not known previously (for instance, in the form of the surface
waveguiding in photonic topological insulators, which is immune to
scattering on defects, as the scattering is suppressed by the topology of
the guiding system), and are unified under the name of \textit{metaoptics}
\cite{metaoptics}. Another unifying concept is \textit{nanophotonics}, the
name originating from the fact that many of these materials are assembled of
elements with sizes measured on the nanometer scale (which is deeply
subwavelength, in terms of optics). One of fundamentally interesting
subjects of nanophotonics is trapping and transmission of light in \textit{%
nanowires}, i.e., optical filaments (usually, made of silicon), whose
diameter, measured in nanometers, is much smaller than the wavelength of
light, while a typical length may be a few millimeters; one of their
important applications is the use in solar photovoltaic elements \cite{wire}.

\subsection{Dissipative and parity-time symmetric waveguides}

The brief discussion of the waveguiding mechanisms give above did not
address the presence of losses and the necessity to compensate them by gain.
This assumption is valid for relatively short propagation distances, as well
as in the case when the compensating gain matches the action of losses so
accurately that both factors may be simultaneously neglected, in the first
approximation. In reality, losses are an inevitably existing gradient in
plasmonics and metamaterials, as the respective waveguides are based on
metallic elements, which introduce the Ohmic dissipation. ]

Generally speaking, if the medium is essentially lossy, the above-mentioned
\textit{index-guiding} and \textit{bandgap-guiding} mechanisms, which define
the guiding channel(s), respectively, in terms of a transverse profile of
the local refractive index, or the transmission-band structure, induced by
the PhC or PhC fiber, may be replaced by a \textit{gain-guiding scheme}, in
which the signal propagates, in a lossy planar or bulk medium, along a
narrow stripe of gain locally embedded into the medium \cite%
{in-book,invited,Olga}.

A recently developed topic, closely related to the light transmission in
dissipative waveguides, deals with \textit{the parity-time }($\mathcal{PT}$)
\textit{symmetry}, which implies balance between symmetrically (in space)
placed gain and loss elements. A paradigmatic model (it often includes
nonlinearity, although the $\mathcal{PT}$ symmetry is, by itself, a linear
property) is represented by NLSE (\ref{NLSE}), in which the potential is
made \textit{complex}, with real and imaginary parts being, respectively,
spatially even and odd ones:
\begin{gather}
i\frac{\partial u}{\partial z}+\frac{1}{2}\frac{\partial ^{2}u}{\partial
x^{2}}-\left[ U_{r}(x)+iU_{i}(x)\right] u+\sigma |u|^{2}u=0,  \notag \\
U_{r}(-x)=U_{r}(x),~U_{i}(-x\_=-U_{i}(x).  \label{PT}
\end{gather}%
Another fundamental realization of the $\mathcal{PT}$ symmetry in optics and
related fields is offered by a coupler, in which one core carries uniformly
distributed gain, and the parallel-coupled one is uniformly lossy, the
accordingly modified Eqs. (\ref{coupler}) being%
\begin{eqnarray}
i\frac{\partial u}{\partial z}+\frac{1}{2}\frac{\partial ^{2}u}{\partial
x^{2}}+\kappa v-U(x)u+\sigma |u|^{2}u &=&i\gamma u,  \notag \\
&&  \label{gamma} \\
i\frac{\partial v}{\partial z}+\frac{1}{2}\frac{\partial ^{2}v}{\partial
x^{2}}+\kappa u-U(x)v+\sigma |v|^{2}v &=&-i\gamma v,  \notag
\end{eqnarray}%
where $\gamma >$ 0 is the gain-loss coefficient. The $\mathcal{PT}$ symmetry
has been experimentally realized in photonics, and a large number of
guided-wave-propagation regimes have were investigated in such systems \cite%
{PT0,PT,PT1,PT2}. In particular, as concerns solitons, although $\mathcal{PT}
$-symmetric systems belong to the class of dissipative ones, where,
generally speaking, solitons exist as isolated \textit{attractors}, selected
by the condition of the double balance, between the dispersion (or
diffraction) and nonlinearity, and between the gain and loss (the latter
principle is very important for the creation of stable temporal solitons in
fiber lasers \cite{Akhm}), in $\mathcal{PT}$-symmetric systems solitons
exists in \textit{continuous families}, similar to their counterparts in
conservative models \cite{PT1,PT2}. In addition to the interest to
fundamental studies, systems with the $\mathcal{PT}$ symmetry offer
promising applications, such as, in particular, \textquotedblleft light
diodes", admitting unidirectional propagation of light in the waveguide, and
lasers operating in the $\mathcal{PT}$-symmetric regime \cite{uni}.

\section{Annotation of articles included in the Special Issue}

The present Special Issue is composed as a collection of \textbf{20}
contributions, which include \textbf{5} relatively brief reviews,
summarizing recently obtained results in various areas of the guided-wave
propagation in photonics, and \textbf{15} original papers reporting novel
findings in this broad field. The contributions may be naturally grouped
according to different forms and manifestations of the guided-wave
propagation addressed in these works. Accordingly, in the list of papers
published in the Special Issue, following below, is divided in 11 topics (%
\textbf{(A)}-\textbf{(K)}, and review articles are highlighted. In all the
cases, subjects addressed in the papers are sufficiently clearly defined by
their titles.

\textbf{(A)} A batch of three papers may be classified as addressing
problems arising in the fundamental (general) theory of the guided wave
transmission in conservative (i.e., lossless) nonlinear media.

\textbf{(A1)} J. Fujioka, A. G\'{o}mez-Rodr\'{\i}guez, and \'{A}.
Espinosa-Cer\'{o}n, \textit{Pulse Propagation Models with Bands of Forbidden
Frequencies or Forbidden Wavenumbers: A Consequence of Abandoning the Slowly
Varying Envelope Approximation and Taking into Account Higher-Order
Dispersion}, Appl. Sci. \textbf{7}, 340 (2017);

\textbf{(A2)} H. N. Chan and K. W. Chow, \textit{Rogue Wave Modes for the
Coupled Nonlinear Schr\"{o}dinger System with Three Components: A
Computational Study},\textit{\ }Appl. Sci. 7, 559 (2017);

\textbf{(A3)} A. Govindarajan, B. A. Malomed, A. Mahalingam, and A.
Uthayakumar, \textit{Modulational Instability in Linearly Coupled Asymmetric
Dual-Core Fibers}, Appl. Sci. \textbf{7}, 645 (2017).

\textbf{(B)} A related topic is the study of bright and dark soliton in
various settings. This topic is represented in the Special Issue by the
following four contributions, one of them being a review article:

\textbf{(B1) }Z. Mai, H. Xu, F. Lin, Y. Liu, S. Fu, and Y. Li, \textit{Dark
Solitons and Grey Solitons in Waveguide Arrays with Long-Range Linear
Coupling Effects}, Appl. Sci. \textbf{7}, 311 (2017);

\textbf{(B2) }G. C. Katsimiga, J. Stockhofe, P. G. Kevrekidis, and Peter
Schmelcher,\textit{\ Stability and Dynamics of Dark-Bright Soliton Bound
States Away from the Integrable Limit}, Appl. Sci. \textbf{7}, 388 (2017);

\textbf{(B3) }P. Rodriguez, J. Jimenez, T. Guillet, and T. Ackemann, \textit{%
Polarization Properties of Laser Solitons}, Appl. Sci. \textbf{7}, 442
(2017);

\textbf{(B4) Review}: F. Mitschke, C. Mahnke, and A. Hause, \textit{Soliton
Content of Fiber-Optic Light Pulses}, Appl. Sci. \textbf{7}, 635 (2017).

\textbf{(C)} Specific aspects of transmission in optical waveguides are
considered in the following three papers (the first two address problems of
direct relevance to practical applications):

\textbf{(C1) }M. Lamy, C. Finot, J. Fatome, J. Arocas, J.-C. Weeber, and K.
Hammani, \textit{Demonstration of High-Speed Optical Transmission at 2
$\mu$%
m in Titanium Dioxide Waveguides}, Appl. Sci. \textbf{7}, 63 (2017);

\textbf{(C2) }F. A. Memon, F. Morichetti, and A. Melloni, \textit{%
Waveguiding Light into Silicon Oxycarbide}, Appl. Sci. \textbf{7}, 561
(2017);

\textbf{(C3) }J. D. Huerta Morales and B. M. Rodr\'{\i}guez-Lara, \textit{%
Photon Propagation through Linearly Active Dimers}, Appl. Sci. \textbf{7},
587 (2017).

\textbf{(D)} Different aspects of the transmission of light in waveguides
based on fiber Bragg gratings is considered in two papers:

\textbf{(D1) }S.-C. Yang, Y.-J. He, and Y.-J. Wun, \textit{Designing a Novel
High-Performance FBG-OADM Based on Finite Element and Eigenmode Expansion
Methods}, Appl. Sci. \textbf{7}, 44 (2017);

\textbf{(D2) Review}: Y. Liu, S. Fu, B. A. Malomed, I. C. Khoo, and J. Zhou,
\textit{Ultrafast Optical Signal Processing with Bragg Structures}, Appl.
Sci. \textbf{7,} 556 (2017).

\textbf{(E)} A specific phenomenon of bound states existing in the
continuous spectrum of a waveguide built as an array of dielectric spheres
is summarized in the following \textbf{Review} article:

E. N. Bulgakov, A. F. Sadreev, and D. N. Maksimov, \textit{Light Trapping
above the Light Cone in One-Dimensional Arrays of Dielectric Spheres}, Appl.
Sci. \textbf{7}, 147 (2017).

\textbf{(F) }A topic of the propagation of self-accelerating beams in the
form of Airy waves is overviewed in a \textbf{Brief Review}:

Y. Zhang, H. Zhong, M. R. Beli\'{c}, and Y. Zhang, \textit{Guided
Self-Accelerating Airy Beams-A Mini-Review}, Appl. Sci. \textbf{7}, 34
(2017).

\textbf{(G)} A specific aspect of the light propagation in metamaterials is
considered in:

V. Mazzone, J. S. Totero Gongora and A. Fratalocchi, \textit{Near-Field
Coupling and Mode Competition in Multiple Anapole Systems}, Appl. Sci.
\textbf{7}, 542 (2017).

\textbf{(H) }Some fundamental aspects of the light transmission in
dissipative waveguides are addressed in the following paper:

O. Descalzi and C. Cartes, \textit{Stochastic and Higher-Order Effects on
Exploding Pulses}, Appl. Sci. \textbf{7}, 887 (2017).

\textbf{(I)} Theoretical studies of the propagation of light in $\mathcal{PT}
$-symmetric nonlinear waveguides are represented by an original paper,

J. D'Ambroise and P. G. Kevrekidis,\textit{\ Existence, Stability and
Dynamics of Nonlinear Modes in a 2D Partially }$\mathcal{PT}$\textit{\
Symmetric Potential}, Appl. Sci. \textbf{7}, 223 (2017).

\textbf{(J)} The propagation of plasmonic waves is addressed in the
following two experimental works, with direct implications for applications:

\textbf{(J1) }K. Moon, T.-W. Lee, Y. J. Lee, and S.-H. Kwon, \textit{A
Metal-Insulator-Metal Deep Subwavelength Cavity Based on Cutoff Frequency
Modulation},

Appl. Sci. \textbf{7}, 86 (2017);

\textbf{(J2)} M. Iwanaga, \textit{Perfect Light Absorbers Made of
Tungsten-Ceramic Membranes}, Appl. Sci. \textbf{7}, 458 (2017).

\textbf{(K)} Specific aspects of the general topic of fiber lasers, which
are significant to fundamental and applied studies alike, are the subject of
a \textbf{Review} article:

C. B. de Ara\'{u}jo, A. S. L. Gomes, and E. P. Raposo, \textit{L\'{e}vy
Statistics and the Glassy Behavior of Light in Random Fiber Lasers}, Appl.
Sci. \textbf{7}, 644 (2017).

\end{document}